\documentstyle[aps,prl,twocolumn]{revtex}

\newcommand{\abv}[1]{\left|{}#1\right|}
\newcommand{\be}[1]{\begin{equation}\label{#1}}
\newcommand{\bbt}{{\cal B}}

\newcommand{\ccc}{{\bf C}}
\newcommand{\cdt}{,\ldots,}
 
\newcommand{\dff}{\sc }

\newcommand{\ha}{h}
\newcommand{\hhh}{{\cal H}}
\newcommand{\ket}[1]{{\bf |}#1{\bf \rangle}}
\newcommand{\kkk}{{\cal K}}
\newcommand{\leadelem}{\ha^{\emptyset}}
\newcommand{\lth}{l}
\newcommand{\lv}[1]{V(#1)}
\newcommand{\lvo}{{V_0}}

\newcommand{\norm}[1]{{\left||{}#1\right||}}
\newcommand{\rf}[1]{(\ref{#1})}
\newcommand{\tbt}{m}

\newcommand{\vvv}{{\cal V}}
\newcommand{\cardinality}{{\rm card}}
\newcommand{\lspan}{{\rm span}}

\newcommand{\ke}{\ket{1}}
\newcommand{\ko}{\ket{0}}

\newcommand{\eprf}{\hspace*{\fill}$\Box$}
\newcommand{\proof}{\noindent {\bf Proof. }}

\newtheorem{dfn}{Definition}
\newtheorem{theorem}{Theorem}
\newtheorem{lemma}[theorem]{Lemma}

\title{
Decomposition of pure states of a quantum register
}

\author{Ioannis Raptis, \\{\em Department of Mathematics, 
University of Pretoria, Pretoria 0002,Republic of South Africa; 
e-mail:  iraptis@math.up.ac.za} \\Roman R. Zapatrin\\{\em Quantum 
Information Group, ISI, Villa Gualino, V.le S.Severo 65,10133, 
Torino, Italy; e-mail:  zapatrin@isiosf.isi.it (address for 
correspondence)}}

\begin{document}

\maketitle

\begin{abstract}
Using the leading vector method, we show that any vector 
$\ha\in(\ccc^2)^{\otimes \lth}$ can be decomposed as a sum of at 
most (and at least in the generic case) $2^\lth-\lth$ product 
vectors using local bitwise unitary transformations. The method is 
based on representing the vectors by chains of appropriate 
simplicial complex. 
This generalizes the Scmidt decomposition of pure states of a 2-bit 
register to registers of arbitrary length $\lth$.  
\end{abstract}

\pacs{PACS Nos. 03.67.-a, 03.65.Bz, 03.65.Fd}

Briefly, the contents of the paper is the following. In any 
computational basis a vector in the register's state space is a sum 
of $2^\lth$ vectors. Each basis vector in the computational basis 
we associate with a simplex whose dimension is the number of $1$'s 
in its binary label minus one. So, any vector becomes a chain. 
Then we show that by appropriate local unitary transformations we 
can always make the 0-dimensional component of the chain equal to 
zero.  Therefore the resulting chain will contain at most 
$2^\lth-\lth$ terms. 

To make the account self-consistent, begin with necessary 
definitions. Let $\bbt=\ccc^2$, consider a register of $\lth$ bits, 
then its state space is 

\[ 
\hhh=\bbt \otimes \bbt \otimes\cdots\otimes \bbt = \ccc^{2^\lth} 
\] 

\noindent A state $\ha\in \hhh$ is said to be a {\dff product state} 
if it can be decomposed into a product: 

\[ 
\ha=
\ha_1\otimes \ha_2 \otimes\cdots\otimes \ha_\lth 
\] 

\noindent Fix a basis $\{\ko,\ke\}$ in 
$\bbt$, then any basis vector in the product space can be encoded 
as a binary string of $\lth$ components, and $\ha\in \hhh$ can be 
decomposed as: 

\be{edech} 
\begin{array}{rcl}
\ha &=&
\ha^{00\ldots0}\cdot\ko\ko\cdots\ko+
\ha^{10\ldots0}\cdot\ke\ko\cdots\ko+\cdots
\cr&+&\cdots
\ha^{11\ldots1}\cdot\ke\ke\cdots\ke
\end{array}
\end{equation}

\section{Simplicial complexes and chains} 

Let $\vvv$ be a non-empty finite set, call the elements of $\vvv$ 
{\dff vertices}.

\begin{dfn}
A collection $\kkk$ of non-empty subsets of $\vvv$ is called (abstract)
{\dff simplicial complex} with the set of vertices $\vvv$ whenever

\begin{itemize}
\item $\forall v\in\vvv \quad \{v\}\in\kkk$
\item $\forall P\in\kkk,\, \forall Q\subseteq\vvv \quad
Q\subseteq P\Rightarrow Q\in\kkk$
\end{itemize}

\noindent The elements $P\in\kkk$ are called {\dff simplices}.

\end{dfn}

\medskip 

Suppose we have enumerated the vertices of $\kkk$, then any simplex 
of $\kkk$ can be encoded as a binary string. Conversely, any binary 
string corresponds to a subset of vertices. In the special case 
when $\kkk$ is the complex of all faces of a simplex $S$, there is 
1--1 correspondence between all binary strings and simplices of 
$\kkk$. The zero string $00\cdt0$ is associated with the empty 
simplex $\emptyset$. 

\noindent {\bf Chains.} The {\dff dimension} of a simplex $P$ is the 
number of its vertices minus one:

\be{e53}
\dim P = \cardinality P - 1
\end{equation}

\noindent Denote by $\kkk^n$ the $n$-skeleton of $\kkk$ --- the 
set of its simplices of dimension $n$

\[
\kkk^n =
\{P\in\kkk:\:\dim P = n\}
\]

\noindent and consider the linear spans

\[
\hhh^n = \lspan\kkk^n =
\left\{\sum_{P\in\kkk^n}c_P\cdot\ket{P}\right\}
\]

\noindent The elements of $\hhh^n$ are called {\dff chains} of 
dimension $n$.  The direct sum 

\be{ecc} 
\hhh=
\oplus \hhh^n
\end{equation}

\noindent is called the {\dff complex of chains} of $\kkk$. The 
decomposition \rf{ecc} endows $\hhh$ with the structure of graded 
linear space. 

\medskip 

\noindent {\bf Simplicial representation of the register space.} We 
shall represent the states of the $N$-bit register by chains of the 
appropriate simplicial complex. Consider a simplex $S$ whose 
vertices are in 1--1 correspondence with the bits of the register, 
and let $\kkk$ be the complex of all faces of $S$, including the 
empty one. Consider the decomposition \rf{edech} of an arbitrary 
state of the register. We see that the basic elements are in 1--1 
correspondence with the simplices of $\kkk$, that is, we can 
consider the vectors of the register's state space as chains of 
the complex $\hhh$ and write down \rf{edech} as the following sum

\be{esimpbas}
\ha=
\sum_{s\in\kkk}\limits\ha^s\ket{s}
\end{equation}

\medskip 

\begin{dfn} 
Let $s,t$ be two simplices and $v$ be a vertex of 
$S$ such that $v\in s$ and $v\not\in t$. The following expression 
will be called {\dff exchangeability condition} of a chain $\ha$: 

\be{exc} 
\ha^s\cdot \ha^t\;=\;\ha^{s\setminus v}\cdot \ha^{t\cup v}
\end{equation} 

\end{dfn} 

\medskip 

The following lemma gives us a necessary and sufficient condition 
for $\ha$ to be a product vector. 

\begin{lemma}\label{lweak} 
A vector $\ha\in\hhh$ is product if and only if for any $s,t$ the 
exchangeability condition \rf{exc} holds. 
\end{lemma} 

\proof 
With no loss of generality we can assume that $\leadelem\neq0$ (it 
can always be achieved by appropriate 
swapping of labelling basis vectors by 0 and 1). Let us reconstruct 
the factors giving the product. The overall phase factor will be 
the phase factor of $\leadelem$. Take it out, then $\leadelem$ 
becomes apositive real number. Then normalize the vector: 
$\ha\mapsto\frac{\ha}{\norm{\ha}}$. Reconstruct the parameter 
$\alpha_1$ from: 

\[ 
\tan\alpha_1=
\left|\frac{\ha^{10\ldots00}}{\leadelem}\right| 
\] 

Then reconstruct the phase $e^{i\phi_1}$ setting it equal to that 
of $\ha^{10\ldots00}$. And repeat this consecutively for all other 
vertices (=bits of the register). 

Now we have to prove that the vector 

\[
\lv{\ha}=
\otimes_{k=1}^\lth
(\cos\alpha_k\ko+e^{i\phi_1}\sin\alpha_k\ke) 
\] 

\noindent is that what we have started with. Denote every simplex 
$s$ by $s=\sigma^1\sigma^2\ldots\sigma^n$, each 
$\sigma=0$ or 1. Then calculate 

\be{ehb} 
\lv{\ha}^s= 
\prod_{k=1}^\lth\limits\; 
\left(
(1-\sigma_k)\cos\alpha_k\ko+
\sigma_k\exp{i\phi_k}\sin\alpha_k\ke
\right)
\end{equation}  

\medskip 

We have to prove now that $\ha^s=\lv{\ha}^s$ for any simplex 
$s\in\kkk$.  For $s$ whose binary string contains at most one 1, 
this is so by construction. For $s$ of dimension 1 ({\em i.e.} 
containing 2 vertices $s=\{u,v\}$ it is so due to \rf{exc}. When we 
have proved it for for all 1-dimensional simplices, the 
exchangeability condition \rf{exc} allows us to prove it for all 
2-dimensional simplices, and so on. 

\eprf

\section{The leading vector}

In this section we present a way to extract the greatest product 
component from a given vector $\ha\in\hhh$. Given a basis $\kkk$ in 
$\hhh$, decompose $\ha$:

\[ 
\ha=\sum_{s\in\kkk}\limits\ha^s\ket{s}
\] 

\noindent and form the vector $\lvo\ha$: 

\[ 
\lvo\ha=\otimes_{k=1}^\lth
(\ha^{\emptyset}\ko+\ha^k\ke)
\] 

\begin{lemma} If $\ha$ is a product state such that 
$\leadelem\neq0$ then 

\[ 
a=\frac{\mbox{
$\lvo\ha$}}{\mbox{
$(\leadelem)^{\lth-1}$}}
\] 

\end{lemma}

\proof 
Is verified by checking the exchangeability condition \rf{exc}. 
\eprf 

\begin{dfn}
Let $\ha$ be a vector in $\hhh$ such that $\leadelem\neq0$ with 
respect to a given computational basis. Then the vector $\lv{\ha}$ 
is called a {\dff leading vector} of $\ha$: 

\be{edeflv}
\lv{\ha}=
\frac{\mbox{
$\lvo\ha$}}{\mbox{
$(\leadelem)^{\lth-1}$}}=
\frac{\mbox{
$1$}}{\mbox{
$(\leadelem)^{\lth-1}$}}\cdot
\otimes_{k=1}^\lth
(\ha^{\emptyset}\ko+\ha^{\{k\}}\ke)
\end{equation} 

\end{dfn}

\noindent Note that the mapping $\ha\mapsto\lv{\ha}$ is only 
uniform rather than linear: 

\[
\lv{\lambda\ha}=\lambda\lv{\ha}
\] 

\begin{lemma}\label{lnumterdec}
Let $\ha\in\hhh$ and $\leadelem\neq0$ in a given basis $\vvv$. Then 
the decomposition of the residual vector $\ha'=\ha-\lv{\ha}$ with 
respect to the basis $\kkk$ \rf{esimpbas} contains at most 
$2^\lth-\lth-1$ nonzero terms.  
\end{lemma} 

\proof 
It follows directly from the formula \rf{edeflv} that 

\[ 
\begin{array}{lcl}
(\lv{\ha})^\emptyset
&=&
(\ha)^\emptyset
\cr
(\lv{\ha})^{\{k\}}
&=&
(\ha)^{\{k\}}\quad\mbox{for any $k=1,\ldots,\lth$}
\end{array}
\] 

\noindent therefore the residual vector $\ha'=\ha-\lv{\ha}$ will 
contain at least $\lth+1$ zero terms when decomposed in the basis 
$\kkk$.
\eprf

\noindent {\bf The leading vector after local transformations.} Suppose 
we have made a local transformation $\ha\mapsto U_\tbt\ha$ in the 
$\tbt$-th bit, then the squared norm of the leading vector of the 
transformed $\ha$ is: 

\be{esqnorm}
\begin{array}{rcl}
\kappa_\ha(U_\tbt)&=&
\norm{\lv{U_\tbt\ha}}^2=
\cr&=&
\abv{\frac{1}{(U_\tbt\ha^\emptyset)^{\lth-1}}}^2
\prod_{k=1}^\lth\limits
(\abv{U_\tbt\ha^{\emptyset}}^2+\abv{U_\tbt\ha^{\{k\}}}^2)
\end{array} 
\end{equation}

\medskip 

\noindent In a similar way we can define the the value 
$\kappa_\ha(U)$ for any $U=\otimes_\tbt{}U_\tbt$. For any fixed 
vector $\ha$ the dependence $U\mapsto\kappa_\ha(U)$ is a continuous 
function on a compact set $\otimes_\tbt{}SU(2)$, therefore it takes 
its maximal value for a particular 
$U=\otimes_\tbt{}U_\tbt\in\otimes_\tbt{}SU(2)$. Choose the new 
basis $\kkk'$ making transformations in each $\tbt$-th bit 

\be{enewbit}
\left(
 \begin{array}{c}
 \ket{0'}
 \cr
 \ket{1'}
 \end{array}
\right)
\;=\;
U_\tbt
\left(
 \begin{array}{c}
 \ko
 \cr
 \ke
 \end{array}
\right)
\end{equation}
 
\medskip 

\begin{lemma}
In the basis $\kkk'$ \rf{enewbit} 

\[
\lv{\ha}=\leadelem\ket{\emptyset}
\] 

\end{lemma}

\proof 
Consider an infinitesimal local transformation $U_\tbt(\tau)$ at 
$\tbt$-th bit. Due to \rf{esqnorm} the norm of the derivative is 
proportional to the value of $\ha^{\{\tbt\}}$. Since $\kkk'$ is 
chosen so that it is maximal, all the derivatives should be zero, 
therefore $\ha^{\{\tbt\}}=0$ for any $\lth=1\cdt \lth$.  
\eprf 

\noindent {\bf Corollary.} In the basis $\kkk'$ \rf{enewbit} 

\[
\lv{\ha}\,\perp\,(\ha-\lv{\ha}
\] 

\noindent therefore 

\be{eorthdcmp}
\ha=\lv{\ha}\,\oplus\,(\ha-\lv{\ha}
\end{equation} 

\medskip 

According to lemma \ref{lnumterdec} the second summand contains at 
most $2^\lth-\lth-1$ terms. Therefore we have decomposed $\ha$ into 
$2^\lth-\lth$ orthogonal product states. In the generic case the 
number of terms is equal to $2^\lth-\lth$ since any infinitesimal 
local transformation of the basis $\kkk'$ increases the number of 
terms in the decomposition. 

\section*{Summary}

The proposed techniques generalize the well-known Schmidt 
decomposition of vectors in the state space of a bipartite system 
to the case of tensor product of any number $\lth$ of 2-dimensional 
systems. Schematically, the decomposition looks as follows. 

{\em Given a vector $\ha\in\otimes_k\ccc^2$, choose a local unitary 
transformation which, applied to the basis, maximizes the component 
$\ha^{00\cdt0}$. }

According to lemma \ref{lnumterdec}, $\ha$ will be decomposed into 
at most $2^\lth-\lth$ orthogonal product states. In particular, for 
$\lth=2$ this means that any bipartite pure state is in generally a 
sum of 2 product states (Schmidt decomposition), any 3-partite pure 
state decays into at most 5 product states, 4-partite pure state is 
a sum of at most 12 product states and so on.

\section*{Acknowledgments}

The work was carried out under the auspices of the European 
research project IST-Q-ACTA. IR thanks prof. Anastasios Mallios 
(University of Athens) for numerous technical exchanges, RRZ 
expresses his gratitude to the ISI quantum computation research 
group, in particular to Christof Zalka, for the attention to the 
work and helpful advice. 

\end{document}